\documentclass[12pt]{article}
\usepackage[left=2cm, right=2cm, bottom=2cm ,top = 2cm, footskip=1cm]{geometry}
\usepackage[font=small,labelfont=bf]{caption}
\usepackage{multirow}
\usepackage{setspace}
\usepackage{tabularx}
\usepackage{lineno}

\let\counterwithin\relax
\usepackage{chngcntr}
\usepackage{float}
\usepackage{hyperref}
\usepackage{tcolorbox}
\usepackage[authoryear,sort&compress]{natbib}
\usepackage{natbib}
\setcitestyle{open={(},close={)}}
\usepackage{adjustbox}
\usepackage{graphicx}
\usepackage{amsmath}

\usepackage[section]{placeins}

\title{Combined impact of grey and superficial white matter abnormalities: implications for epilepsy surgery}
\author{Csaba Kozma$^{1*}$, Jonathan Horsley$^{1}$, Gerard Hall$^{1}$, Callum Simpson$^{1}$, Jane de Tisi$^{3}$,\\  Anna Miserocchi$^{3}$, Andrew W. McEvoy$^3$, Sjoerd B. Vos$^{3,4,5}$, Gavin P. Winston$^{3,6}$, \\ Yujiang Wang$^{1,2,3}$, John S. Duncan$^{3}$,  Peter N. Taylor$^{1,2,3}$}

\begin{document}
\maketitle
\begin{singlespace}
% author affiliations
\begin{enumerate}
\item{CNNP Lab (www.cnnp-lab.com), School of Computing, Newcastle University, Newcastle upon Tyne, United Kingdom}
\item{Translational and Clinical Research Institute, Faculty of Medical Sciences, Newcastle University, Newcastle upon Tyne, United Kingdom}
\item{Department of Clinical and Experimental Epilepsy, UCL Queen Square Institute of Neurology, University College London, London, United Kingdom}
\item{Western Australia National Imaging Facility, The University of Western Australia, Nedlands, Australia}
\item{Centre for Medical Image Computing, Computer Science Department, University College London, London, United Kingdom}
\item{Department of Medicine, Queen's University, Kingston, Ontario, Canada}

\end{enumerate}

\begin{center}
* c.a.kozma2@newcastle.ac.uk   
\end{center}

%Highlights
\footnotesize

HIGHLIGHTS:
\begin{itemize}
  
  \item Abnormalities in both grey matter and superficial white matter are found in epilepsy patients, helping inform surgical decisions.
  \item Incorporating both GM and SWM together improves surgical outcome differentiation.
  \item Removing concordant abnormal regions further increases the chance of seizure freedom.
\end{itemize}


We confirm that we have read the Journal’s position on issues involved in ethical publication and affirm that this report is consistent with these guidelines. Supporting information is included in a separate file.
\end{singlespace}

\newpage

\section*{Abstract}

Drug-resistant focal epilepsy is associated with abnormalities in the brain in both grey matter (GM) and superficial white matter (SWM). However, it is unknown if both types of abnormalities are important in supporting seizures. Here, we test if surgical removal of GM and/or SWM abnormalities relates to post-surgical seizure outcome in people with temporal lobe epilepsy (TLE).

We analyzed structural imaging data from 143 TLE patients (pre-op dMRI and pre-op T1-weighted MRI) and 97 healthy controls. We calculated GM volume abnormalities and SWM mean diffusivity abnormalities and evaluated if their surgical removal distinguished seizure outcome groups post-surgically.

At a group level, GM and SWM abnormalities were most common in the ipsilateral temporal lobe and hippocampus in people with TLE. Analyzing both modalities together, compared to in isolation, improved surgical outcome discrimination (GM AUC = 0.68, p $<$ 0.01, WM AUC = 0.65, p $<$ 0.01; Union AUC = 0.72, p $<$ 0.01, Concordance AUC = 0.64, p = 0.04). Additionally, 100\% of people who had all concordant abnormal regions resected had ILAE$_{1,2}$ outcomes.

These findings suggest that regions identified as abnormal from both diffusion-weighted and T1-weighted MRIs are involved in the epileptogenic network and that resection of both types of abnormalities may enhance the chances of living without disabling seizures.

\newpage
\section{Introduction}

Approximately half of the people who undergo epilepsy surgery still experience seizures postoperatively, likely due to incomplete resection of the epileptogenic zone \citep{deTisi2011}. Successful outcomes depend on the precise localization and removal or disconnection of seizure-generating tissue \citep{Duncan2016, Bell2017}. Although grey matter (GM) and superficial white matter (SWM) abnormalities are often assessed separately, they may jointly contribute to the epileptogenic network. This study examines the relationship between resection of both GM and SWM abnormalities and post-surgical outcomes in individuals with temporal lobe epilepsy (TLE).

People with TLE commonly have GM atrophy in the temporal, frontal and parietal cortices, along with subcortical structures, including the thalamus and amygdala, and damage to WM tracts, especially in the limbic and subcorticocortical regions, particularly close to the epileptogenic zone \citep{Larivière2020, Whelan2018, Liu2016, Hatton2020}. GM atrophy in TLE often follows a widespread, multi-lobar, bilateral pattern. WM damage is predominately lateralized to the hemisphere of seizure focus \citep{Larivière2020, Galovic2019}. Hippocampal sclerosis (HS) is associated with both SWM damage and extensive cortical thinning \citep{Liu2016, Bonilha2013, Mueller2009}. SWM abnormalities can complement T1w imaging by revealing epileptogenic zone features invisible on T1w alone \citep{Liu2016, Schilling2023}. Few studies have specifically investigated the connection between GM and WM abnormalities \citep{Winston2020, Chen2024, Horsley2022}, and cortical thinning and damage to WM may result from separate underlying processes \citep{Liu2016, Winston2020}.

Resection of structurally abnormal regions is associated with an increased chance of seizure freedom \citep{Galovic2019, Bonilha2007, Fadaie2024, Bernhardt2015b, Borger2021}. Furthermore, removing nodes or tracts with abnormal diffusion may also improve outcomes \citep{Horsley2024, Sinha2021, Keller2017, Bonilha2015}. SWM improves tissue characterization by detecting subtle abnormalities and precisely localizing epileptic regions, guiding surgical decisions to optimize patient outcomes \citep{Horsley2024}. Although each approach may aid in identifying epileptogenic tissue, few studies have examined GM and WM abnormalities together in the context of epilepsy surgery \citep{Liu2016, Yasuda2010}. Our study aims to fill this gap and emphasize the advantages of jointly analysing GM and SWM abnormalities in this context. 

This study examines the relationship between the resection of abnormal regions in GM and SWM and post-surgical outcomes in 143 individuals with TLE who underwent resective surgery. We identified regions characterized by reduced GM volume and increased SWM mean diffusivity (MD) and compared these regions to individual resection masks. This allowed us to evaluate how effectively each modality — individually and in combination — differentiates between favorable outcomes (ILAE$_{1,2}$) and less favorable ones (ILAE$_{3+}$).

\section{Methods}

\subsection{Subjects}

We conducted a retrospective study of 143 individuals with temporal lobe epilepsy (TLE) at the National Hospital for Neurology and Neurosurgery, London, who had temporal lobe resections between 2009 and 2021. This study of anonymized data that had previously been acquired was approved by the Health Research Authority, without the need to obtain consent from each subject (UCLH epilepsy surgery database: 22/SC/0016), and by the Database Local Data Monitoring Committee. Individuals who declined for their data to be used in anonymised research were not included in the research database. Patients were matched with 97 healthy controls, who provided individual written consent, (Table 1), and data were collected using two acquisition protocols. All participants underwent anatomical T1-weighted and diffusion-weighted MRI, with TLE laterality assessed through pre-surgical evaluations. Age at onset of epilepsy ranged from 1 to 52 years (median = 15 years, IQR = 15.75 years) and 53\% of patients had HS (Table 1). Postoperatively, 73.4\% achieved ILAE$_{1,2}$ seizure outcomes at 12 months (Table 1). Results regarding ILAE outcomes for Years 2–5 are provided in the Supplementary Material.

\captionof{table}{ Control and Patient data by outcome at 1 yr} \label{tab:title}
\begin{center}
\begin{tabular}{ |p{5cm}|p{2.1cm}|p{2.1cm}|p{2.1cm}|p{4.0cm}|} 
\hline
 & Control & $ILAE_{1,2}$  & $ILAE_{3+}$ & Test statistic \\
\hline
N & 97 & 105 & 38 &  \\ 
Age of Onset, median (IQR) & - & 15 (15) & 17.5 (11.8) &  $W = 2381, p = 0.08$ \\ 
Age at Scan, median (IQR) & 39 (20.8) & 36.9 (17.3) & 41.7 (15.4) &  $K = 2.409, p = 0.30$ \\ 
Sex, male:female & 37:60 & 43:62 & 19:19  & $X^{2} = 1.32, p = 0.52$ \\
Side, left:right & - & 62:43 & 19:19  & $X^{2} = 0.60, p = 0.43$ \\
HS, Yes:No & - & 60:45 & 16:22 &  $X^{2} = 9.49, p = 0.15$ \\
\hline
\end{tabular}
\end{center}

\subsection{Data acquisition}
Individuals were scanned using one of two acquisitions. The first cohort (87 patients, 29 controls) was scanned between 2009 and 2013 using a 3T GE Signa HDx scanner with standard imaging gradients (40 mTm-1, 150 Tm-1s-1) and an 8-channel phased array coil. T1-weighted images were acquired with a 3D inversion-recovery fast spoiled gradient recalled echo (FSPGR) sequence (TE= 3.04ms, TR= 37.68s, 170 contiguous 1.1 mm coronal slices each a 256 x 256 matrix with 0.9375 x 0.9375 mm in plane resolution). Diffusion-weighted images (DWI) were collected with a cardiac-triggered single shot spin-echo planar imaging (EPI) sequence (TE= 73 ms) with 60 axial slices. Each slice was 2.4mm in size with a 96 x 96 matrix, zero-filled to 128 x 128, with 1.875 x 1.875mm in-plane resolution. Overall 52 diffusion directions (b = 1,200 s/mm$^2$ [$\delta$=21Ms, $\Delta$=29ms, maximum gradient strength]) with 6 B0 scans were collected.

\vspace{5mm}

The second cohort (56 patients, 68 controls) was scanned between 2014 and 2019 using a 3T GE MR750 scanner with a higher gradient strength (50 mTm-1, 200 Tm-1s-1) and a 32-channel phased array coil. T1-weighted images were collected with a FSPGR sequence (TE= 3.1ms, TR= 7.4ms, TI=400 ms, 170 contiguous 1 mm coronal slices each 256 x 256 with 1mm x 1mm in plane resolution). Diffusion-weighted MRI used a cardiac-triggered single-shot EPI sequence (TE= 74.1ms) with 70 axial slices and 115 volumes across four b-values (0, 300, 700, 2500 s/mm$^2$ [$\delta$=21.5ms, $\Delta$=35.9ms]). Overall, 11 B0 images were collected, with a field of view of 256 x 256 mm and an acquisition matrix size was 128 x 128. The final reconstructed voxel size was 2 × 2 × 2 mm.

\subsection{Data processing and registration}

The dMRI scans from both cohorts underwent identical data processing, including de-noising \citep{Veraart2016}, Gibbs-unringing \citep{Kellner2016}, and signal drift correction \citep{Vos2017}. As one cohort lacked reverse phase-encoded B0s, the Synb0-DisCo tool \citep{Schilling2019, Schilling2020} generated non-distorted synthetic images from T1 structural MRIs for all participants. This tool was applied to both cohorts to maintain consistency. The synthetic images were then signal bias corrected using N4 \citep{Tustison2010}, and processed with TOPUP \citep{Andersson2003, Smith2004} and EDDY to address warping \citep{Andersson2016}, and eddy current distortions, and motion. After pre-processing, tensor maps were calculated via FSL’s DTIFIT tool \citep{Smith2004}. MD maps were computed directly registered to standard space using linear (affine) and non-linear (SYN-Diffeomorphic) registrations using the ANTs toolbox  \citep{Avants2011}. All MD maps were registered to the "FMRIB158\_2mm" MD template provided from the FSL toolbox \citep{Smith2004}. We investigated MD maps as these were recently shown to hold potentially valuable localizing information \citep{Horsley2024}. Once registered, both linear and non-linear transformations were applied to all tensor maps with a trilinear interpolation. To examine SWM, only WM voxels within 5mm of the grey–white matter boundary in the selected regions of interest (ROI) were included in the analysis. A threshold of 5mm was chosen based on previous work demonstrating this as a limit for the largest difference to controls \citep{Hall2025}.

\vspace{5mm}

T1-weighted MRIs were used to generate parcellated GM ROI. FreeSurfer’s ‘recon-all’ pipeline \citep{Fischl2012} was applied for intensity normalization, skull stripping, subcortical volume generation, and parcellation, following ENIGMA guidelines. We used the most detailed version of the Lausanne parcellation \citep{Hagmann2008}, with 446 neocortical and 14 deep brain regions, including the hippocampus, amygdala, thalamus, putamen, and caudate. All FreeSurfer generated surfaces were visually inspected for accuracy and corrected where necessary as described previously \citep{Simpson2025}.

\subsection{Abnormality Calculation}

\begin{figure}[H]
	\centering
	\includegraphics[width=\textwidth]{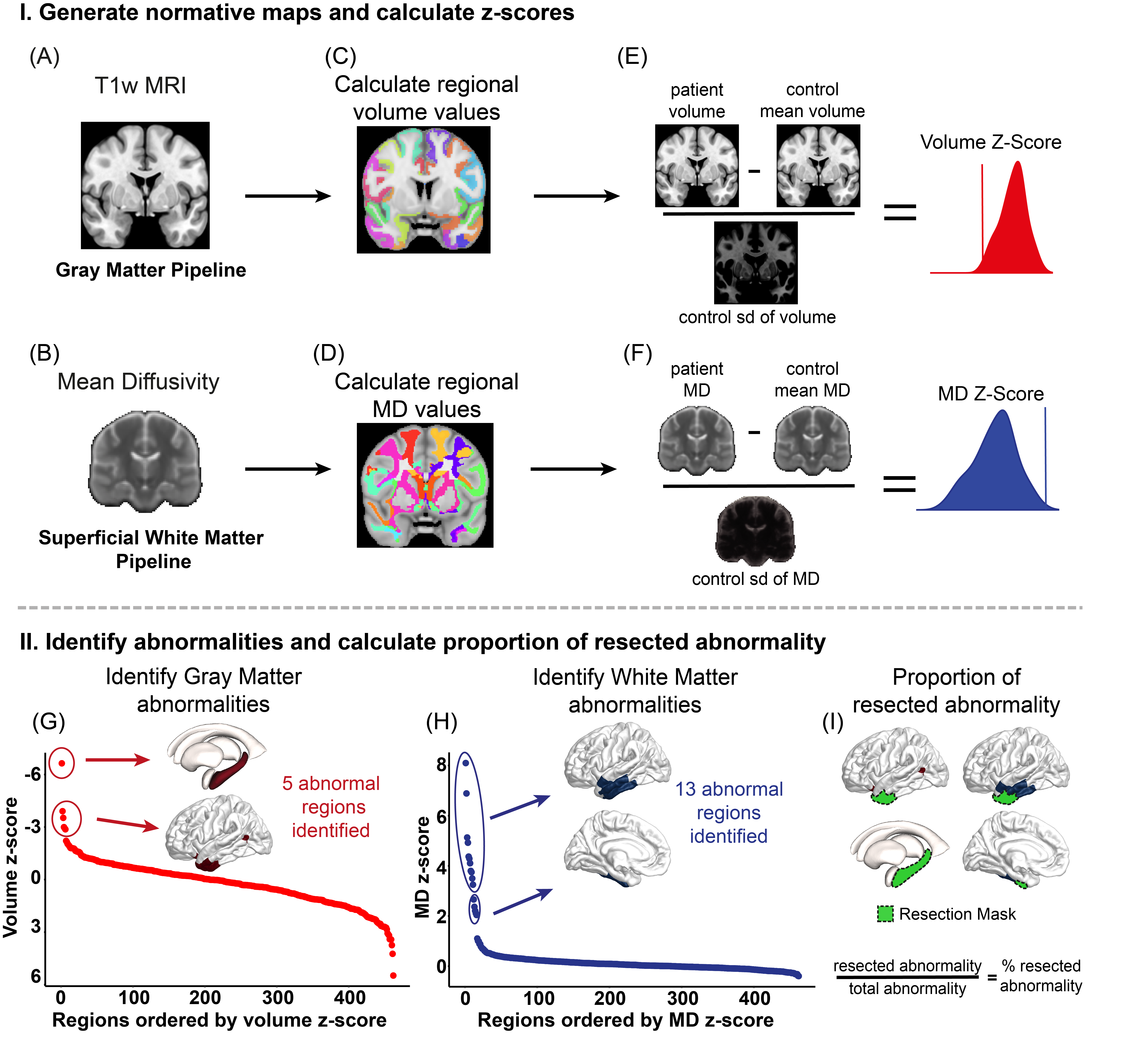}
	\caption{\textbf{I. Generate normative maps and calculate z-scores. } (a, b) Individual T1W MRI and Mean Diffusivity (MD) maps provide volume and superficial white matter MD values. (c, d) Regional volume and MD values are calculated, then (e, f) z-scored against the normative map to indicate deviations from the healthy mean. \textbf{II. Identify abnormalities and calculate proportion of resected abnormality.} (g, h) Z-scores are ranked per modality, and change point analysis identifies abnormal regions (ROIs). (i) For each subject, abnormality distributions are overlaid with the resection mask to calculate the proportion of resected abnormal ROIs.}
	\label{fig:GM_WM abnormalities}
\end{figure}

We used the healthy controls to compute normative baselines of GM volume and adjacent SWM mean diffusivity (MD) per region. We selected MD as it reflects average diffusion in all directions, with higher MD values indicating more unrestricted diffusion, often linked to myelin disruption and increased extracellular space in WM \citep{Arfanakis2002, Concha2009}, which are common in epilepsy \citep{Hatton2020}. For the SWM, we took the mean MD of white matter voxels within 5mm of the nearest region in the Lausanne parcellation. We calculated regional means and standard deviations, harmonizing across scanning protocols using ComBat \citep{Johnson2007} and adjusting for covariates (age and sex). For individuals with TLE, we calculated abnormalities by z-scoring each region’s MD and volume values against the corresponding normative map. These abnormalities quantified deviations from the healthy mean in each region for both modalities (Figure \ref{fig:GM_WM abnormalities}A-F). We focused on negative z-scores for GM abnormalities, reflecting volume reductions. For SWM, we analysed positive z-scores in MD, which indicate increased diffusion \citep{Whelan2018, Hatton2020}. 

Within each patient, regional abnormalities were ranked in each modality separately. Then, multiple change point analysis (MCP) was applied to identify patient-specific thresholds to find abnormal ROIs  (Figure \ref{fig:GM_WM abnormalities}G,H) \citep{Lindeløv2020}. MCP uses Bayesian regression to locate shifts in means, variances, or autocorrelation \citep{Lindeløv2020}; here, it identified mean shifts as thresholds for abnormal ROIs compared to the rest of the distribution of regions. We examined the spatial distribution of regional abnormalities in GM, SWM, their union (GM, SWM or both), and concordance of GM and SWM. We have included a table with these terms and their definitions in the Supplementary material (Table S11).

\subsection{Resection mask generation}

Resection masks were generated using a semi-automated method \citep{Taylor2024}. Postoperative imaging was used to create masks of the resected tissue in preoperative space. This process began with an automated pipeline incorporating FastSurfer \citep{Henschel2020}, ANTs \citep{Avants2011}, and ATROPOS \citep{Avants2011b} to produce the initial masks. These masks were then visually inspected for accuracy and manually corrected if necessary. Once finalized, the resection masks were aligned to the MNI-152 standard space, matching the space of the abnormality maps. Regions with more than 10\% overlap with the resection mask were considered as resected. 

% Methods

For each subject, the abnormality distributions were overlaid with the resection mask to compare the spatial overlap of resected tissue with abnormal ROIs (Figure 1I). We calculated the proportion of abnormal ROIs that had been resected and investigated whether this differed between ILAE$_{1,2}$ and ILAE$_{3+}$ cases.

\subsection{Code and data availability}

Code and data to reproduce figures in the manuscript will be made available upon acceptance of the paper. 

\subsection{Statistical analysis}

After computing the proportion of abnormalities resected in GM, SWM, or both, we compared these values between groups who were subsequently free of disabling seizures (ILAE$_{1,2}$) or not (ILAE$_{3+}$). To measure the magnitude of difference between groups we measured the area under the receiver operating characteristic curve (AUC). To assess its significance, we used a one tailed Kruskal-Wallis test as we had a prior hypothesis that patients without disabling seizures would have a greater proportion of their abnormalities removed.

\section{Results}

\subsection{Abnormality is present in grey and superficial white matter}

Abnormalities in both GM and SWM were present across the whole cohort (Figure 2). GM atrophy in neocortical regions was seen in a widespread, bilateral pattern affecting temporal, frontal and parietal lobes. The greatest proportion of abnormalities were in the ipsilateral hippocampus (47\%), and, with ipsilateral temporal pole 12\% of patients showing abnormally reduced volume (Figure 2A). SWM abnormalities were more discrete, primarily ipsilateral and near to the epileptogenic zone. 26\% had abnormal MD adjacent to the ipsilateral hippocampus, and 25\% of patients had abnormal ipsilateral inferior temporal gyrus MD (Figure 2B).

% GM and SWM abnormality spatial distribution
\begin{figure}[H]
	\centering
	\includegraphics[scale=.95]{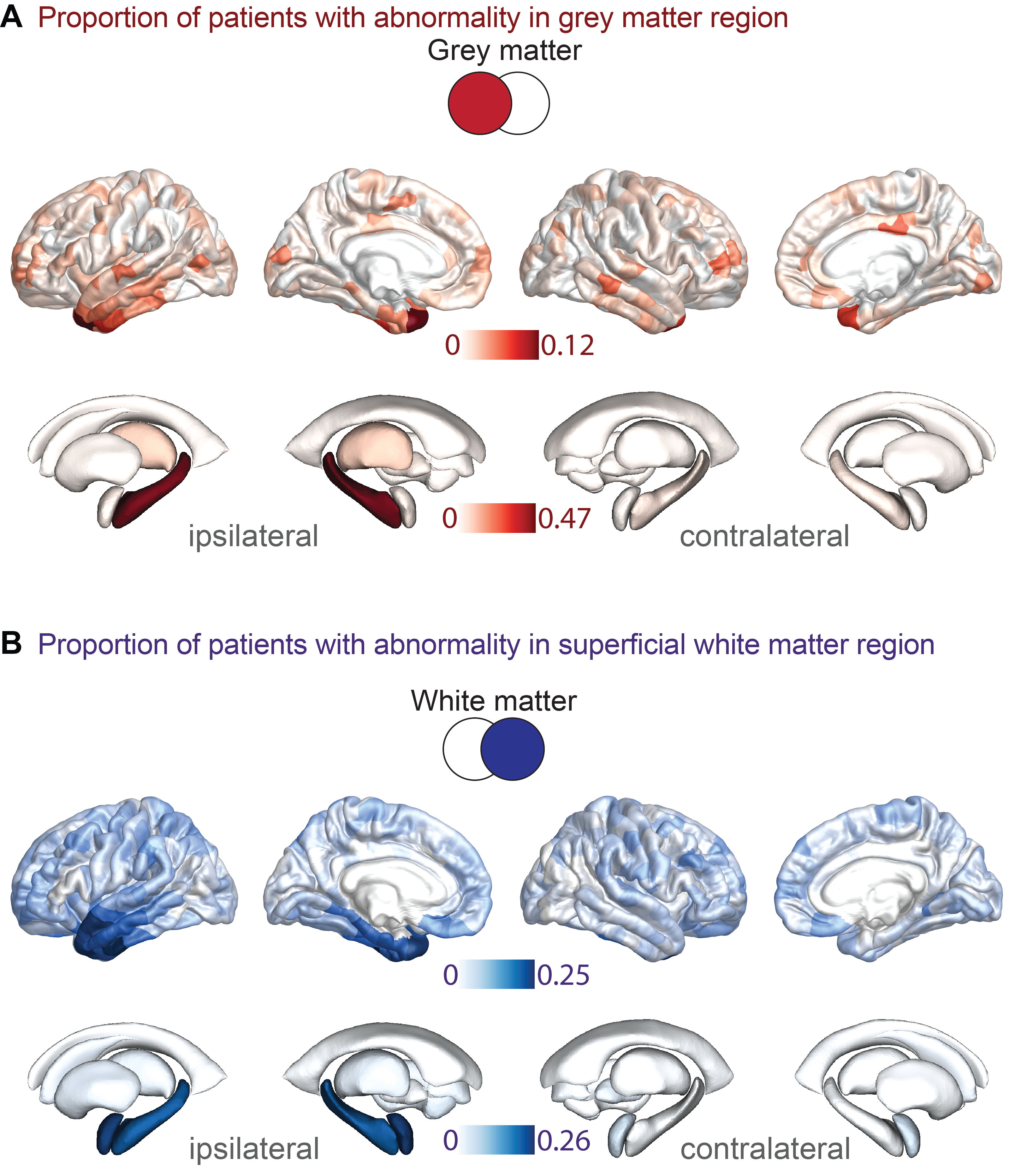}
	\caption{\textbf{Spatial distribution of grey and superficial white matter abnormalities. } The proportion of individuals with abnormalities in each ROI for (A) grey matter and (B) superficial white matter.}
	\label{GM_WM abnormalities}
\end{figure}

\subsection{Resection of grey or superficial white matter abnormalities relates to post-surgical seizure outcome}

Next, abnormality distributions were overlaid with resection masks to calculate the proportion of resected abnormal ROIs and compare ILAE$_{1,2}$ and ILAE$_{3+}$ cases. Resecting abnormal GM ROIs significantly differentiated patient outcomes (AUC = 0.68, p $<$ 0.01; Figure 3A). Similarly, resecting abnormal SWM ROIs significantly differentiated patient outcomes (AUC = 0.65, p $<$ 0.01; Figure 3B). Thus, when considered separately, the resection of GM or SWM abnormalities explains post-surgical seizure outcome moderately well.

% GM and SWM abnormality outcome prediction
\begin{figure}[H]
	\centering
	\includegraphics[scale=1]{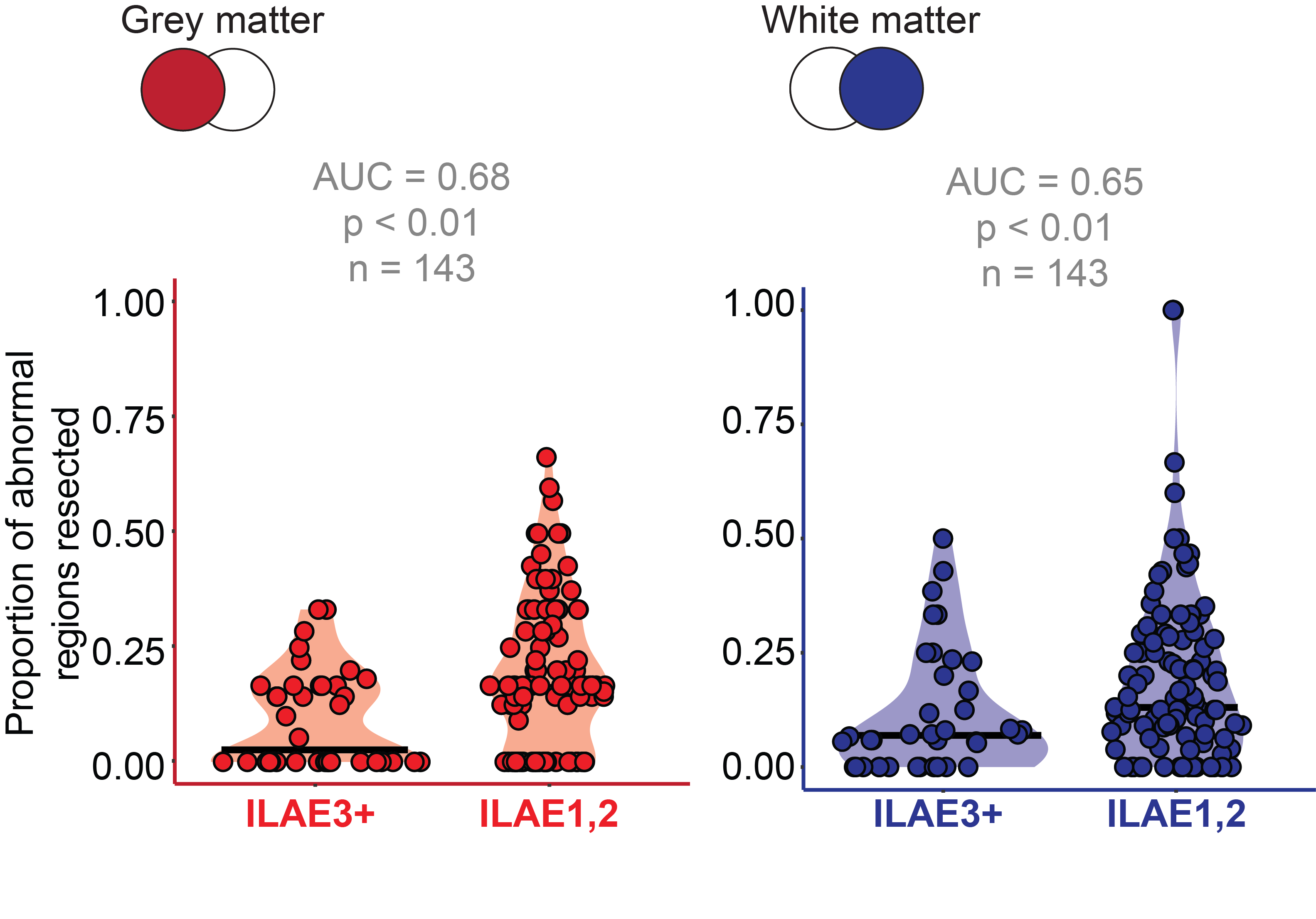}
	\caption{\textbf{: Resecting grey and superficial white matter abnormalities differentiates post-surgical outcome.} The proportion of abnormal regions resected distinguished post-surgical outcome using both (A) GM and (B) SWM. Each data point represents an individual patient. The darker line shows the median. AUCs were derived from ROC curves using the resected ROIs to differentiate surgical outcomes.}
	\label{Outcome_GM_WM}
\end{figure}

\subsection{Combining grey and superficial white matter abnormalities improves outcome distinction}

We investigated whether resection of ROIs containing abnormalities is associated with a greater chance of freedom from disabling seizures, irrespective of their GM or SWM origin (Figure 4AB). The surgical resection of ROIs containing GM, or SWM, or both better differentiated seizure outcome compared to each of the modalities in isolation (AUC = 0.72, p $<$ 0.01; Figure 4A). This result suggests that both modalities separately and together are indicated in abnormal activity.
 \\
Results showed that concordant abnormalities (GM and SWM overlap) were present in 56\% of cases but moderately predicted post-surgical seizure outcome (AUC = 0.64, p = 0.04; Figure 4C). Notably, all patients with complete resection of abnormal concordant ROIs (most frequently the ipsilateral hippocampus) had ILAE$_{1,2}$ outcomes (Figure 4CD).
 
\begin{figure}[H]
	\centering
	\includegraphics[scale=1]{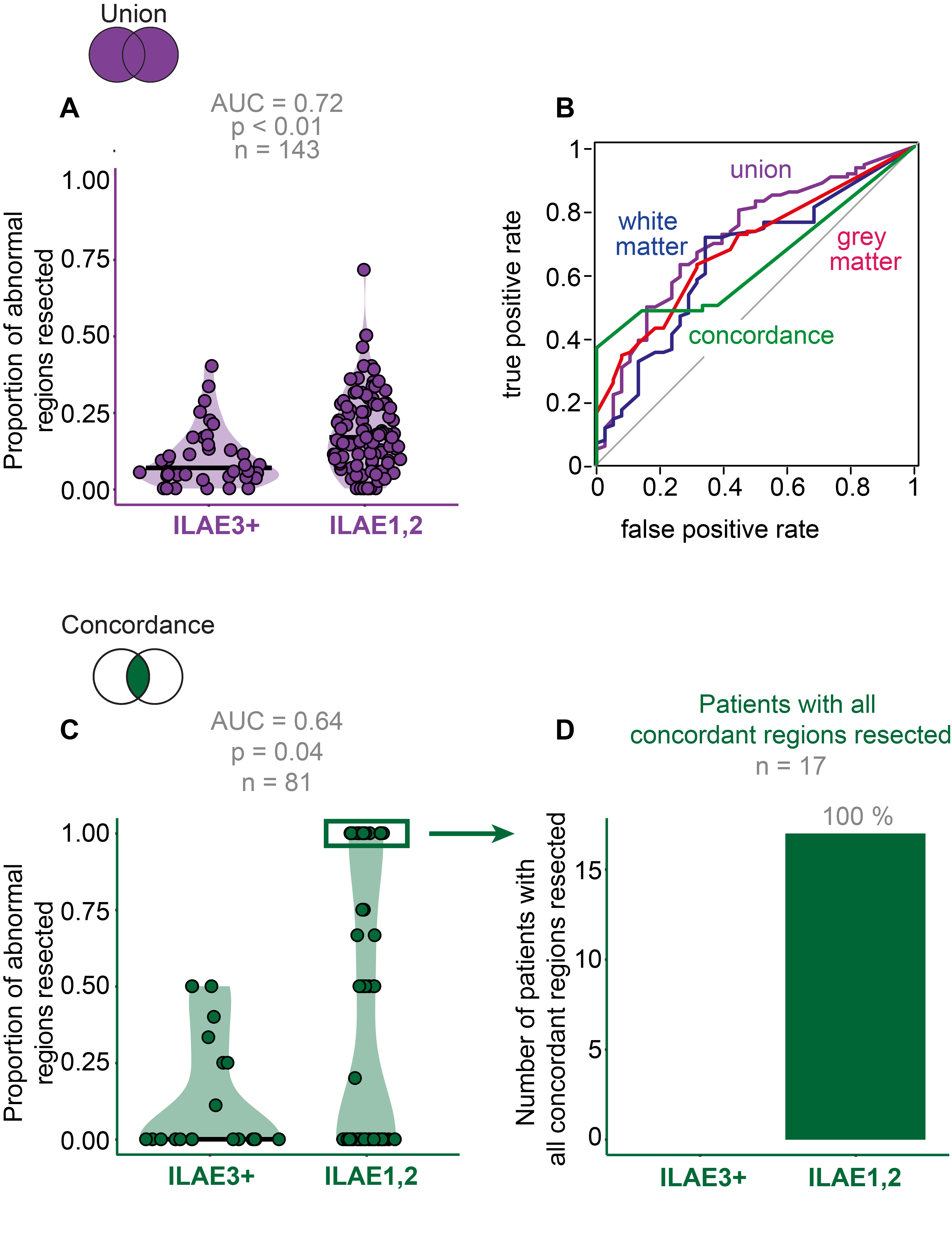}
	\caption{\textbf{Resecting abnormal regions in grey and superficial white matter (union) improves prediction of post-surgical seizure outcome. All patients with complete resection of concordant abnormal ROIs achieved ILAE$_{1,2}$ outcomes.} 
	(A) Proportion of resected abnormal regions (GM and SWM union) distinguishes patients without disabling seizures. Each point represents a patient; the darker line shows the median. 
	(B) ROC curve comparing GM, SWM, their union, and concordance in predicting surgical outcomes. 
	(C) Resected abnormal ROIs compared between ILAE$_{1,2}$ and ILAE$_{3+}$ patients using GM and SWM concordance. Each point represents a patient; the darker line shows the median. 
	(D) Surgical outcomes for patients with all concordant ROIs resected.}
	\label{union abnormalities}
\end{figure}

For completeness, we present discordant GM and SWM abnormalities (Figure S1), which differentiate patient outcomes as well as their union — unsurprising since most union abnormalities are discordant. This supports the value of multimodal analysis. We also show the spatial distribution of union, concordant, and discordant abnormalities (Figure S2). Results remained consistent across acquisition protocols (Figure S4), HS and non-HS cases (Figure S6), and MRI negative cases (Figure S3). Coarser Lausanne parcellations (216+14 and 114+14 regions) confirmed consistent outcome predictions (Figure S5), and surgical outcome findings were replicated across ILAE Year 2–5 outcomes (Figure S7). GM and SWM z-score correlations showed no significant difference between patients and controls (W = 7119, p = 0.6592) (Figure S8). Bootstrapping analysis with 1,000 resamples showed consistent AUCs across the data sets, confirming the reliability of the model (Figure S9). Finally, in the supplementary material (Table S10), we present the average proportions of regions identified as abnormal and the average proportions of abnormal regions resected across the patient cohort for GM, WM, the union, concordance, and discordance of GM and WM.

\section{Discussion}

This study combined GM and SWM abnormalities to identify epileptogenic tissue in TLE in the context of epilepsy surgery. Looking at each modality in isolation, we found the resection of GM or SWM abnormalities was clearly associated with outcome. By combining the two modalities together to include all abnormality, we found group differentiation improved substantially. Finally, while concordance was present in only 56\% of patients, complete resection consistently led to ILAE$_{1,2}$ outcomes.

\vspace{5mm}

This study advances preoperative imaging research for distinguishing TLE surgical outcomes by examining combined GM and SWM abnormalities, which previous studies typically analyzed separately. Removing GM regions with abnormal volume, particularly in the hippocampus and entorhinal cortex, may be beneficial \citep{Galovic2019, Bonilha2007, Fadaie2024, Bernhardt2015b}. We observed GM atrophy in the ipsilateral hippocampus \citep{Whelan2018, Fadaie2024, Keller2015}, extending bilaterally into temporal, frontal, and centroparietal areas \citep{Larivière2020, Galovic2019}. Diffuse GM abnormalities, noted by \citep{Whelan2018, Ballerini2023}, aid in differentiating outcomes, with frontal GM atrophy and hippocampal atrophy or resection extent as reliable indicators \citep{Fadaie2024, Doucet2015}. Our results indicate that GM abnormalities alone more effectively distinguish ILAE$_{1,2}$ from ILAE$_{3+}$ outcomes in HS cases than in non-HS cases, underscoring the importance of hippocampal volumetry in clinical evaluation. SWM abnormalities, concentrated ipsilaterally near the epileptogenic zone \citep{Hatton2020, Urquia-Osorio2022, Chang2019}, also correlate with HS, which associates with cortical thinning \citep{Liu2016, Bonilha2013, Mueller2009}. In TLE, resecting the temporal pole (including the piriform cortex), hippocampus, and adjacent white matter is critical \citep{Sone2022, Dasgupta2023}. Advances in hippocampal mapping show that posterior hippocampal resection is often unnecessary \citep{Dasgupta2023}. Regions such as the ipsilateral hippocampus, parahippocampal gyri, amygdala, and adjacent temporal WM structures show consistent GM/WM abnormalities across studies \citep{Whelan2018, Liu2016}. These mesiotemporal abnormalities often extend to frontal lobes, thalamus, and cingulate cortex, mirroring the diffuse GM atrophy and SWM changes noted in our results. Additionally, anterior temporal lobe resection offers a higher likelihood of seizure freedom compared to lesionectomy \citep{deTisi2011, Horsley2024} found that resecting large SWM abnormality clusters significantly improves seizure freedom. Our findings suggest that targeting both GM and SWM abnormalities together may enhance the likelihood of living without disabling seizures compared to addressing them separately.

\vspace{5mm}

We found that combining GM and SWM abnormalities, considering both their union and discordant contributions, substantially improves surgical outcome differentiations over assessing each in isolation. This suggests that distinct biological processes might drive these two types of injury in TLE, as discussed in prior studies \citep{Liu2016, Winston2020, Chen2024, Ballerini2023, Horsley2022}. The coexistence of GM volume loss and SWM abnormalities in regions such as the temporal pole and frontal SWM underscores their combined role in surgical outcomes, where resection of concordant GM-SWM regions (e.g., anterior hippocampus and adjacent WM) improves the rate of living without disabling seizures. However, our finding that GM abnormalities better predict outcomes in HS cases aligns with prior studies \citep{Whelan2018, Ballerini2023, Fadaie2024} showing HS-associated cortical thinning. Although GM-SWM concordance is uncommon, it appears occasionally, especially in the ipsilateral hippocampus. However, we found that concordance differentiate better in non-HS cases, which does not completely align with findings that highlight concordance in HS cases \citep{Liu2016, Urquia-Osorio2022, Beheshti2018}. This could point to distinct etiologies. While limited in frequency, concordant abnormalities when resected likely lead to ILAE$_{1,2}$ outcomes and should be considered in surgical planning.

\vspace{5mm}

We analyzed GM and SWM abnormalities from a regional perspective, but it is crucial to acknowledge that epilepsy is a network disorder \citep{Bonilha2015, Bernhardt2015, Bernhardt2019, Taylor2018}. Surgical resection can significantly affect the epileptogenic network, potentially preventing further seizures even if regions deemed abnormal are not removed. Prior research demonstrated the impact of surgery on the structural connectome \citep{Bonilha2015, Taylor2018}, with certain regions being more critical due to their higher connectivity.

\vspace{5mm}
Balancing the removal and protection of SWM in epilepsy surgery requires careful consideration. SWM may play a role in the propagation of seizures and the localization of epileptogenic zones \citep{Liu2016, Schilling2023, Ballerini2023}, but modern techniques, such as laser interstitial thermal therapy (LITT) \citep{Ko2022, Bezchlibnyk2018} and historical approaches, such as amygdalohippocampectomy \citep{Gross2018, Ko2022}, prioritize minimizing functional damage. To navigate this balance, a risk-benefit analysis is essential, where decisions are individualized to weigh seizure control against potential risks such as cognitive deficits \citep{Duncan2024, Helmstaedter2013}. Selective resection techniques, such as LITT, can be used to precisely target epileptogenic zones while sparing critical SWM \citep{Ko2022, Bezchlibnyk2018}. Improved localization methods, supported by voxel-based identification of SWM abnormalities \citep{Horsley2024}, help refine SWM mapping. Finally, postoperative monitoring is crucial to assess SWM reorganization and functional recovery over time, guiding rehabilitation efforts accordingly.

\vspace{5mm}

This study has some limitations. We focused on TLE, while future studies could extend this to extra-temporal cases \citep{Liu2016, Chen2024, Doucet2015, Urquia-Osorio2022}. Additionally, only mean diffusivity (MD) was included here, whereas other studies have also used fractional anisotropy, axial diffusivity, and radial diffusivity to assess WM abnormalities in epilepsy \citep{Hatton2020, Winston2020, Urquia-Osorio2022, Slinger2016}. We selected MD as it reflects average diffusion in all directions, with higher MD values indicating more unrestricted diffusion, often linked to myelin disruption and increased extracellular space in WM \citep{Arfanakis2002, Concha2009}, which are common in epilepsy \citep{Hatton2020}. Additionally, we analyzed SWM only up to a 5mm depth, while different depths could be used \citep{Guevara2020}, and future studies could examine multiple depths to determine the most sensitive depth for SWM abnormalities. In addition, we do not have access to histological information, which limits our access to differentiate between Type I and Type II HS \citep{Blumcke2013}. Another limitation is that abnormalities were calculated against two small control cohorts. Although the replicability across the two cohorts gives confidence in our results, future research could benefit from normative diffusion-weighted MRI models combined with T1w-MRI models \citep{Ge2024, Little2024, Bethlehem2022}, trained on larger control samples, for a robust baseline.

\vspace{5mm}

In summary, our study presents a method to investigate the relationship between GM and SWM abnormalities in TLE related to epilepsy surgery. We found that these abnormalities were present across the whole cohort and could provide vital insights for surgical decision-making. Analyzing multiple types of abnormality (here GM and SWM) together substantially enhances differentiations of surgical outcomes, with increased rate of living without disabling seizures when abnormal regions are concordant and resected. Our findings suggest that diffusion-weighted MRI abnormalities complement traditional T1-weighted scans, improving pre-surgical assessments and seizure control, especially in complex cases, thus offering a more accurate prognosis for epilepsy.

\newpage

\begin{singlespace}
\section{Funding}
C.K. is supported by Epilepsy Research Institute UK, P.N.T. and Y.W. are both supported by UKRI Future Leaders Fellowships (MR/T04294X/1, MR/V026569/1). G.P.W. and scan acquisition was supported by the MRC (G0802012, MR/M00841X/1). J.S.D., J.d.T. are supported by the NIHR UCLH/UCL Biomedical Research Centre.

\section{CRediT authorship contribution statement}
Peter N. Taylor: Writing – review \& editing, Supervision, Conceptualization, Funding acquisition. Yujiang Wang: Resources, Conceptualization. Andrew W. McEvoy: Resources. Gerard Hall: Resources, Conceptualization. Callum Simpson: Resources. Jane de Tisi: Resources. John S Duncan: Resources, Conceptualization. Writing – review \& editing. Sjoerd B. Vos: writing – review \& editing, Data curation, Resources Gavin P. Winston: Writing – review \& editing, Investigation, Data curation, Resources, Conceptualization, Funding acquisition. Csaba Kozma: Writing – review \& editing, Writing – original draft, Visualization, Methodology, Formal analysis, Data curation, Conceptualization. Jonathan Horsely: Supervision, Conceptualization, Resources.

\section{Declaration of Competing Interest}
None of the authors has any conflict of interest to disclose.

\section{Acknowledgements}
We thank members of the Computational Neurology, Neuroscience \& Psychiatry Lab (www.cnnp-lab.com) for discussions on the analysis and manuscript. The authors acknowledge the facilities and scientific and technical assistance of the National Imaging Facility, a National Collaborative Research Infrastructure Strategy (NCRIS) capability, at the Centre for Microscopy, Characterization, and Analysis, the University of Western Australia.
\end{singlespace}

\newpage
\bibliography{ref}
\newpage

%%%%%%%%%% %%%%%%%%%% SUPPLEMENTARY %%%%%%%%%% %%%%%%%%%% 
\renewcommand{\thefigure}{S\arabic{figure}}
\setcounter{figure}{0}
\counterwithin{figure}{section}
\counterwithin{table}{section}
\renewcommand\thesection{S\arabic{section}}
\setcounter{section}{0}

\section*{Supplementary}

\section{Discordance between GM and SWM abnormalities}

% Discordance resection distribution
\begin{figure}[H]
	\centering
	\includegraphics[scale=1]{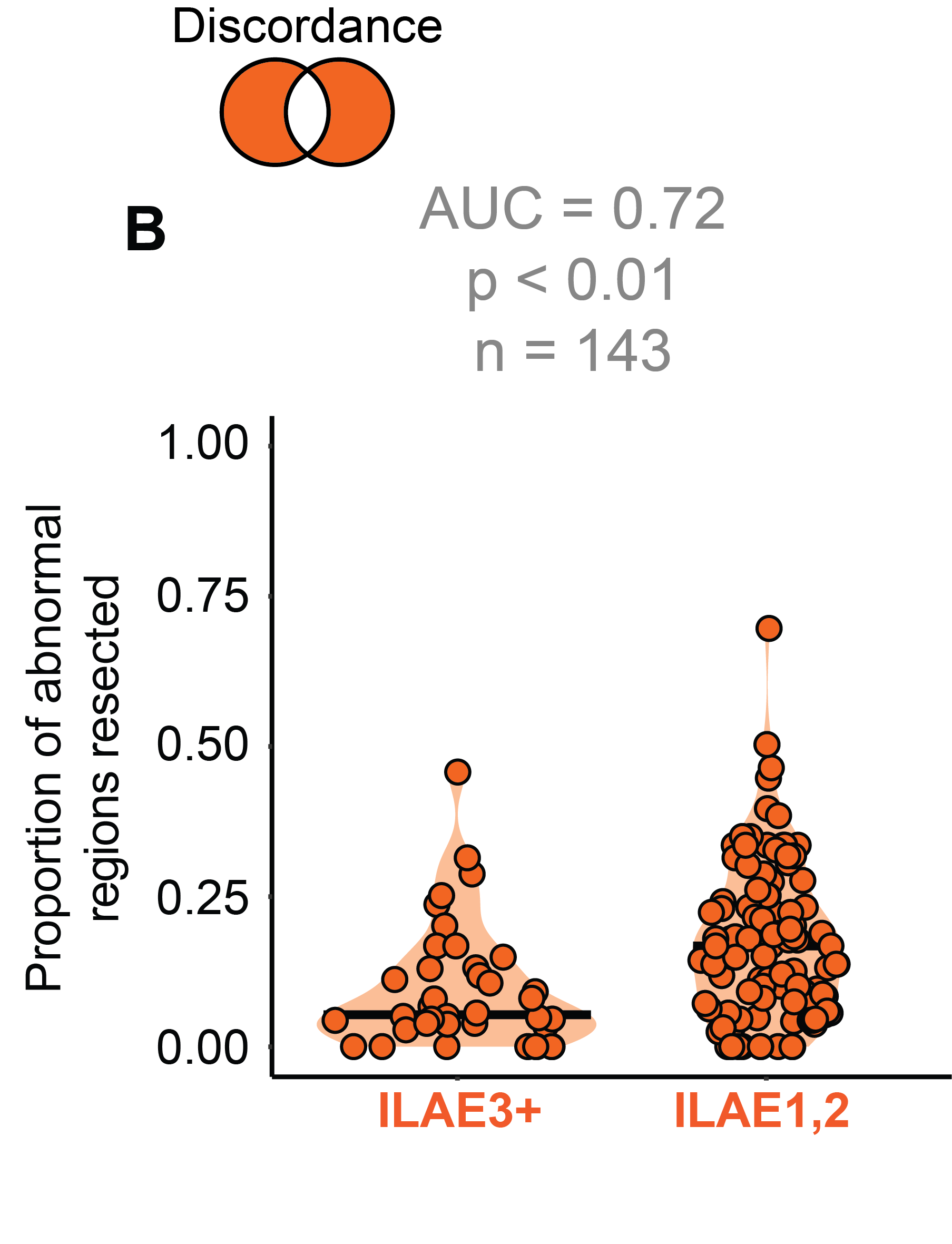}
	\caption{\textbf{Proportion of patients with abnormality in either grey or superficial white matter region (discordance):} (A) Comparison of resected abnormal ROIs between ILAE$_{1,2}$ and ILAE$_{3+}$ patients using GM and SWM discordance. Each point represents a patient, with the darker line indicating the median.}
	\label{disc abnormalities}
\end{figure}

The resection of ROIs containing either GM or SWM but not both (discordant) showed good differentiation of seizure outcomes better than individual modalities (AUC = 0.72, p $<$ 0.01; Figure S1). Discordant and union abnormalities were predominantly located in the ipsilateral hemisphere, particularly near the presumed epileptogenic zone (Figure S2).

\section{Abnormality localization using the combination of GM and SWM: union, concordance, discordance}

% GM/WM combination abnormalities
\begin{figure}[H]
	\centering
	\includegraphics[width=.68\textwidth]{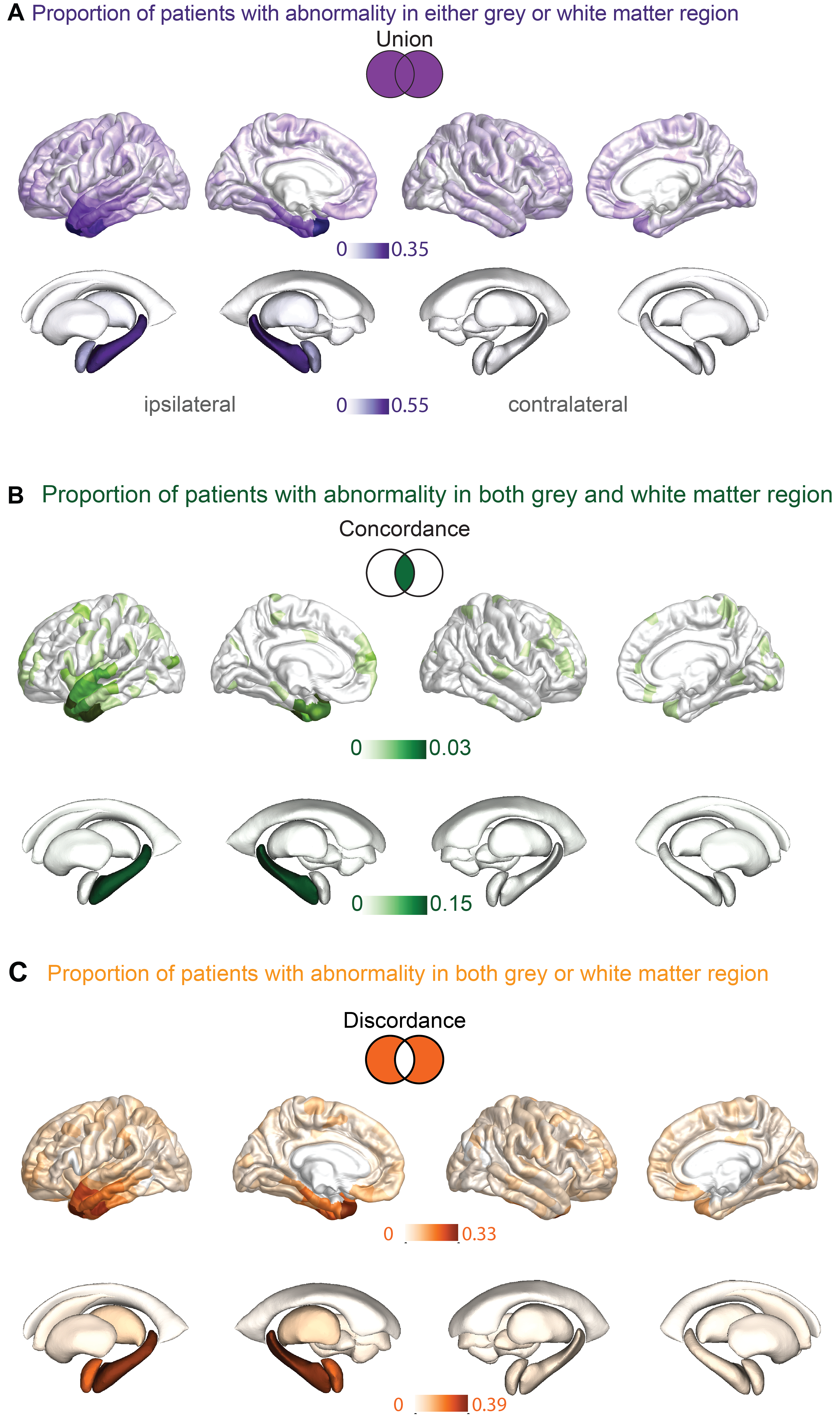}
	\caption{\textbf{Spatial distribution of combination of grey and superficial white matter abnormalities. } The proportion of individuals with abnormalities in each ROI for (A) union, (B) concordance and (C) discordance.}
	\label{GM/WM combination abnormalities }
\end{figure}

 In the union approach, 35\% of patients showed abnormalities in the ipsilateral temporal pole and 55\% in the hippocampus, confirming the hippocampus as the most frequently affected region in the TLE cohort (Figure S2A). The union abnormalities were concentrated in the ipsilateral hemisphere, especially near the presumed epileptogenic zone (Figure S2A). Additionally, concordance between GM and SWM abnormalities occurred only 56\% of cases, with the highest concordance observed in the ipsilateral hippocampus and adjacent SWM (15\%; Figure S2B) and inferior temporal gyrus (3\%). Finally, discordant GM and SWM abnormalities showed similar patterns as union abnormalities with 33\% of patients showing abnormalities in the ipsilateral temporal pole and 39\% in the hippocampus(Figure S2C). These results highlight the impactful contributions of GM and SWM, separately and together, underscoring the potential need for a multimodal approach to epileptic abnormalities.

\section{Differentiate outcome in MRI negative patients}

% MRI negative cases
\begin{figure}[H]
	\centering
	\includegraphics[width=\textwidth]{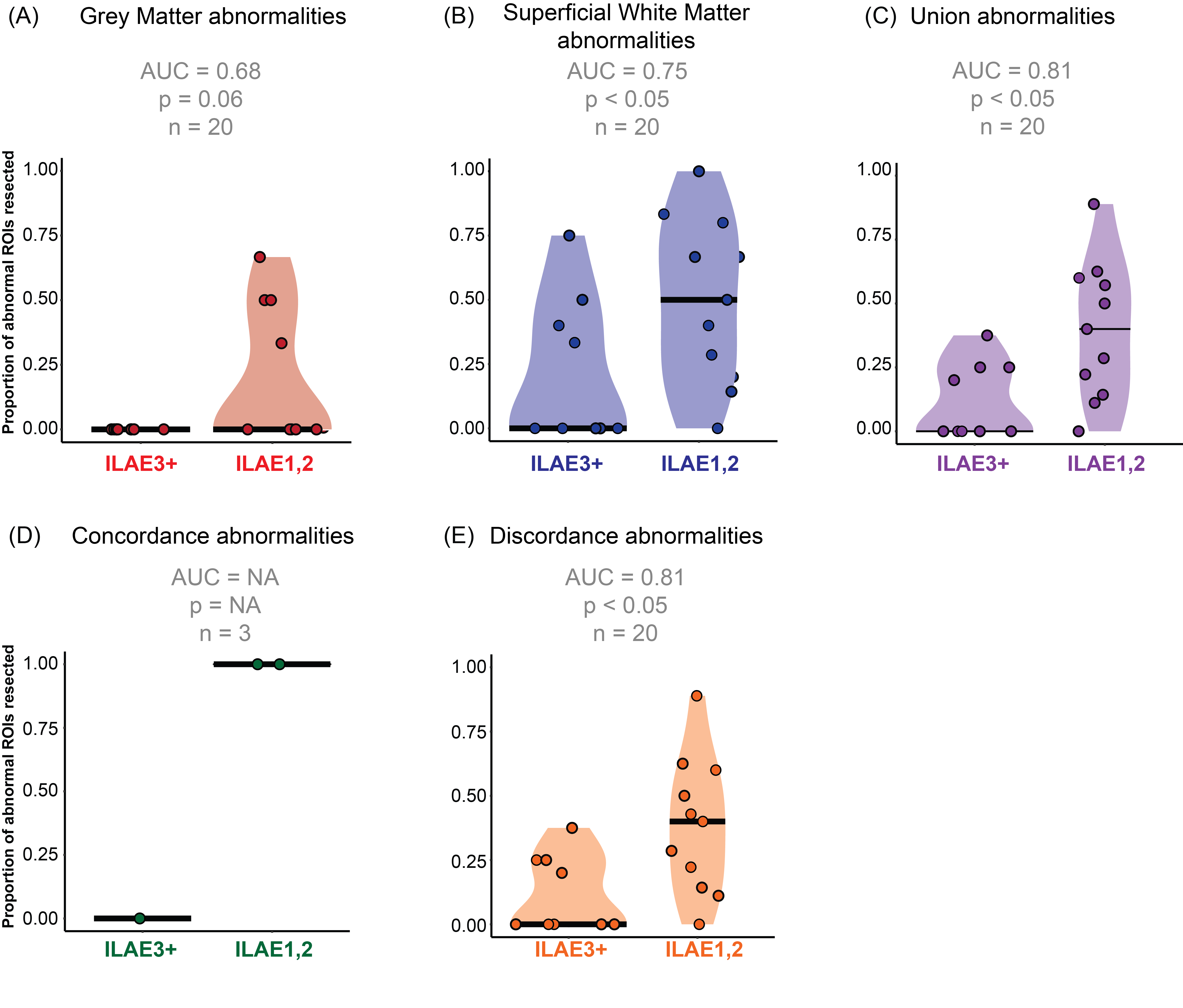}
	\caption{\textbf{Replication of surgical outcome findings in MRI negative individuals using GM, SWM and their combinations: } Comparison of resected abnormal ROIs between ILAE$_{1,2}$ and ILAE$_{3+}$ patients (A) using GM, (B) SWM, (C) union, (D) concordance, (E) and discordance. Each point represents a patient, with the darker line indicating the median.}
	\label{MRI neg abnormalities}
\end{figure}

We replicated our analysis, specifically, in MRI-negative individuals. Of the 143 individuals with TLE, only 20 were MRI-negative, limiting in-depth analysis for this subgroup. Despite the small sample size, here too, union and discordance approaches effectively differentiated ILAE$_{1,2}$ and ILAE$_{3+}$ outcomes in MRI-negative cases (Union AUC = 0.81, p $<$ 0.05; Discordance AUC = 0.81, p $<$ 0.05; Figure S3). These promising findings warrant replication with a larger sample size.

\section{Differentiate outcome separately in our two acquisition cohorts}

% Acq differences
\begin{figure}[H]
	\centering
	\includegraphics[width=\textwidth]{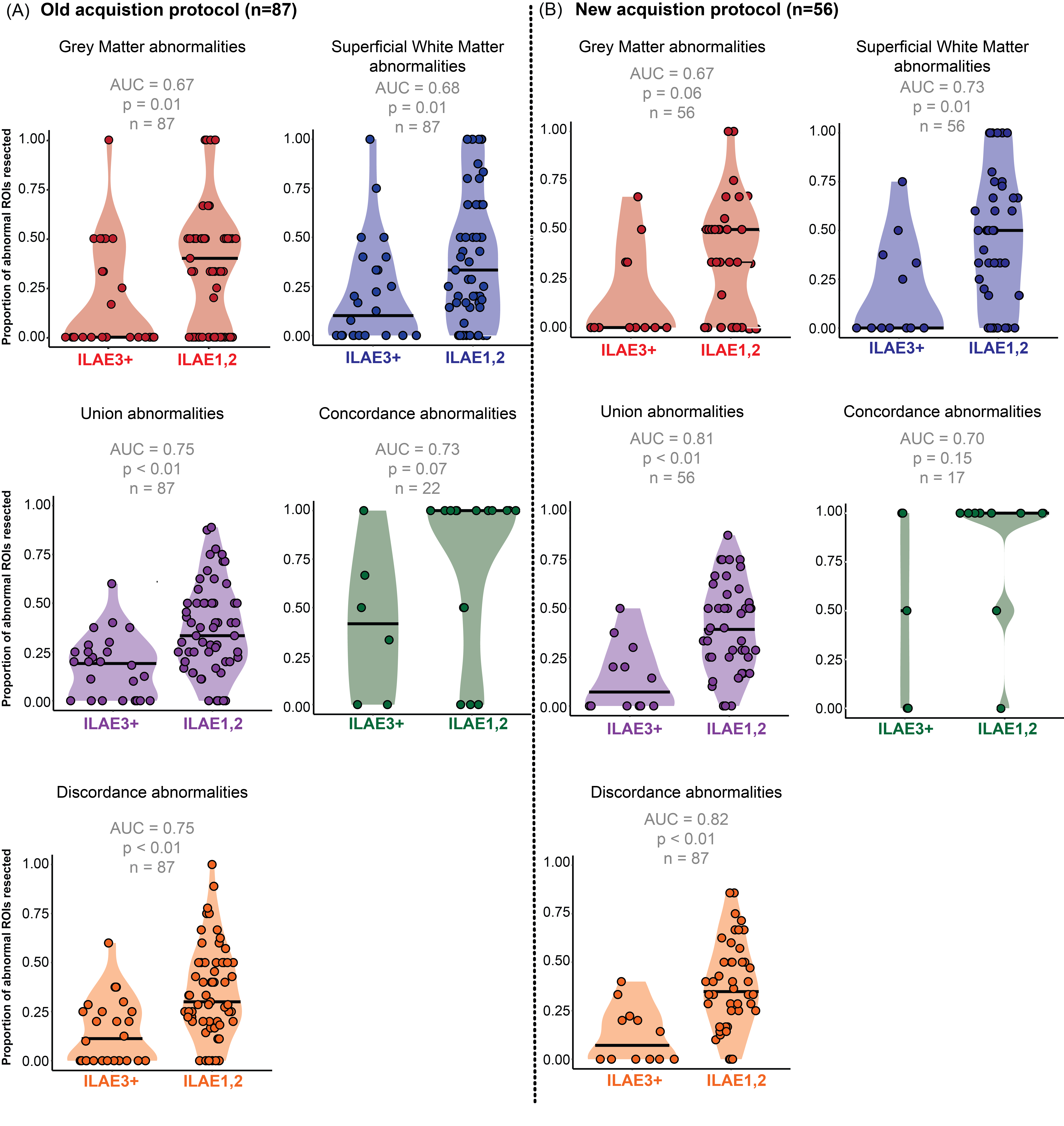}
	\caption{\textbf{Replication of surgical outcome findings in two acquisition cohorts using GM, SWM and their combinations:} Comparison of resected abnormal ROIs in ILAE$_{1,2}$ vs ILAE$_{3+}$ patients using (A) old and (B) new imaging protocols across GM, SWM, union, concordance, and discordance metrics. Each point represents a patient, with a darker line marking the median.}
	\label{MRI neg abnormalities}
\end{figure}

Our data were collected using two acquisition protocols. The first cohort (87 patients, 29 controls) was scanned from 2009 to 2013 on a 3T GE Signa HDx scanner, with T1-weighted images (1.1 mm slices) and DWI using 52 directions (b = 1,200 s/mm²). The second cohort (56 patients, 67 controls) was scanned from 2014 to 2019 on a 3T GE MR750 scanner with improved gradients, T1-weighted images (1 mm slices), and DWI with 115 volumes across four b-values.
Results show that in both cohorts, union and discordance best differentiated ILAE$_{1,2}$ and ILAE$_{3+}$ outcomes (Union old cohort AUC = 0.75, p $<$ 0.01; Discordance old cohort AUC = 0.75, p $<$ 0.01; Union new cohort AUC = 0.81, p $<$ 0.01; Discordance new cohort AUC = 0.82, p $<$ 0.01; Figure S4). The newer acquisition protocol, with improved diffusion quality, outperformed the older protocol, highlighting the importance of updated imaging standards for precise abnormality localization.

\section{Abnormalities across parcellation schemes}
\begin{figure}[H]
	\centering
	\includegraphics[width=\textwidth]{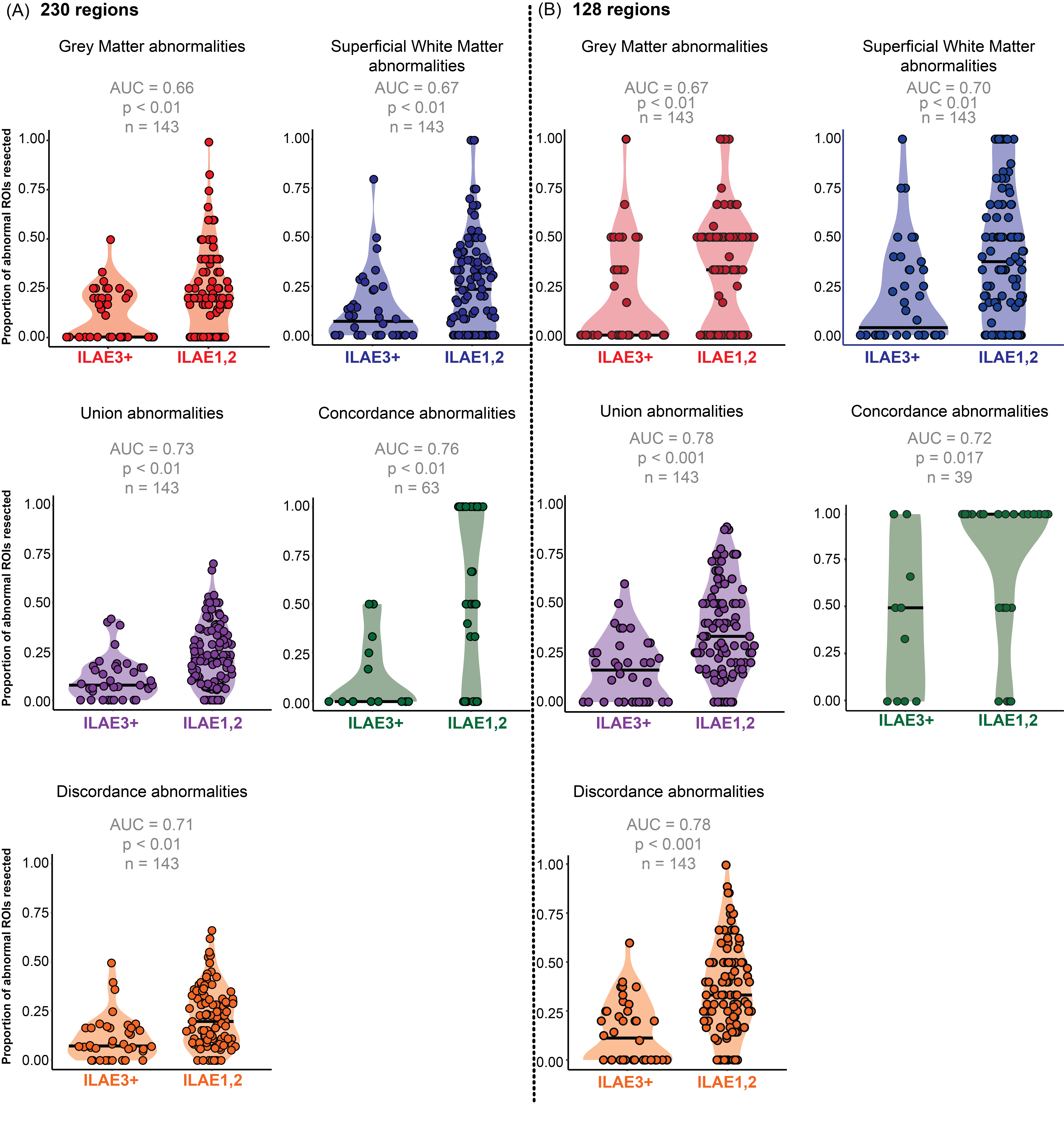}
	\caption{\textbf{Replication of surgical outcome findings across parcellations: } Comparison of resected abnormal ROIs in ILAE$_{1,2}$ and ILAE$_{3+}$ patients in (A) 216 cortical + 14 subcortical and, (B) 114 cortical + 14 subcortical regions across GM, SWM, union, concordance, and discordance metrics. Each point represents a patient, with a darker line marking the median.}
	\label{Parcellation abnormalities}
\end{figure}

We used a 446-region parcellation in the main manuscript. Here, we show consistent results across alternative parcellations (Figure S5). Our results show that union and discordance best differentiate between ILAE$_{1,2}$ and ILAE$_{3+}$ cases across all parcellations.

\section{Abnormalities in HS vs non-HS}
\begin{figure}[H]
	\centering
	\includegraphics[width=\textwidth]{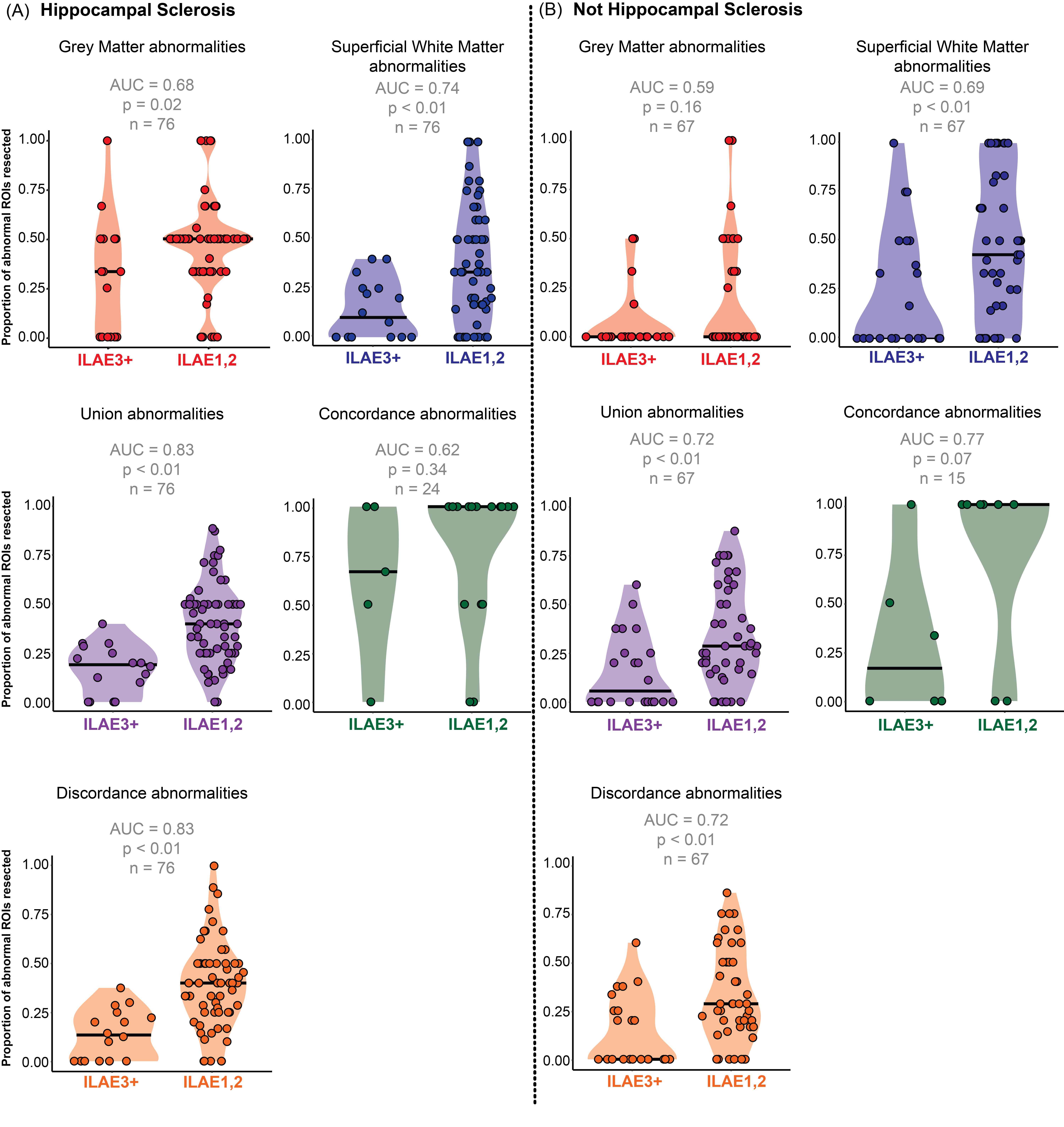}
	\caption{\textbf{Replication of surgical outcome findings in HS and non-HS patients using GM, SWM and their combinations: } Comparison of resected abnormal ROIs in ILAE$_{1,2}$ vs ILAE$_{3+}$ patients splitting to (A) HS and (B) non-HS cases across GM, SWM, union, concordance, and discordance metrics. Each point represents a patient, with a darker line marking the median.}
	\label{MRI HS abnormalities}
\end{figure}

The results held for both HS and non-HS cases, except for GM-only abnormalities (HS AUC = 0.68, p = 0.02; non-HS AUC = 0.59, p = 0.16). GM, SWM, and their combinations better differentiated ILAE$_{1,2}$ and ILAE$_{3+}$ outcomes in HS cases (Figure S6). Surprisingly, concordance differentiated outcomes more effectively in non-HS cases (HS AUC = 0.62, p = 0.34; non-HS AUC = 0.77, p = 0.07), though caution is needed due to the small number of concordant cases.

\section{AUCs for distinguishing ILAE$_{1,2}$ and ILAE$_{3+}$ in Years 2--5.}
\begin{figure}[H]
	\centering
	\includegraphics[width=\textwidth]{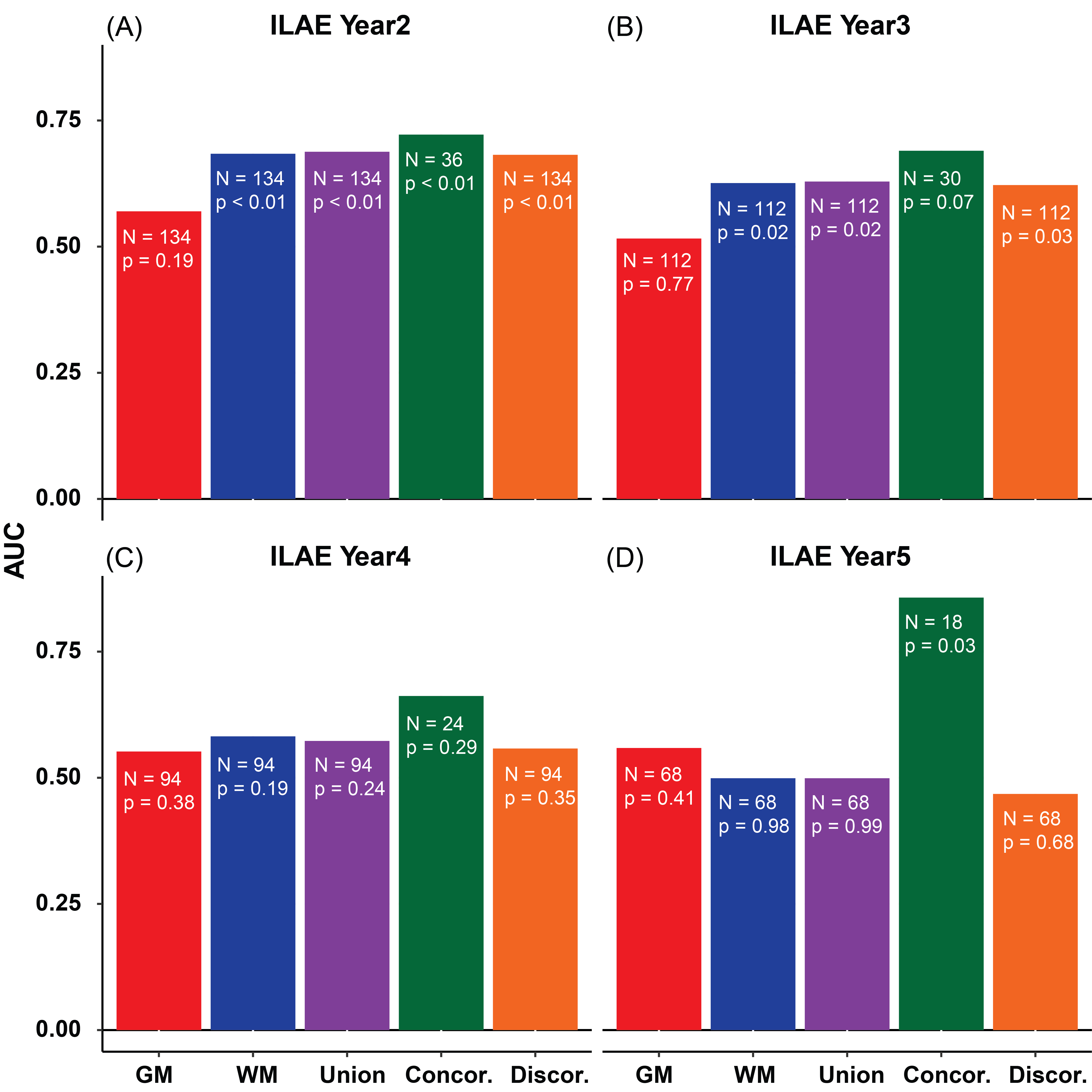}
	\caption{\textbf{Replication of surgical outcome findings across ILAE outcomes Year 2–5: } Comparison of resected abnormal ROIs in ILAE$_{1,2}$ and ILAE$_{3+}$ patients in (A) ILAE Year 2, (B) ILAE Year 3, (C) ILAE Year 4, (D) ILAE Year 5 across GM, SWM, union, concordance, and discordance metrics.}
	\label{ILAEY2-5 abnormalities}
\end{figure}

Postoperatively, we looked at seizure outcomes at 12 months in the main manuscript, while here we present the results for Year 2–5 (Figure S7).

\section{Comparison of GM and SWM z-score correlation: Patients vs. Controls}
\begin{figure}[H]
	\centering
	\includegraphics[width=\textwidth]{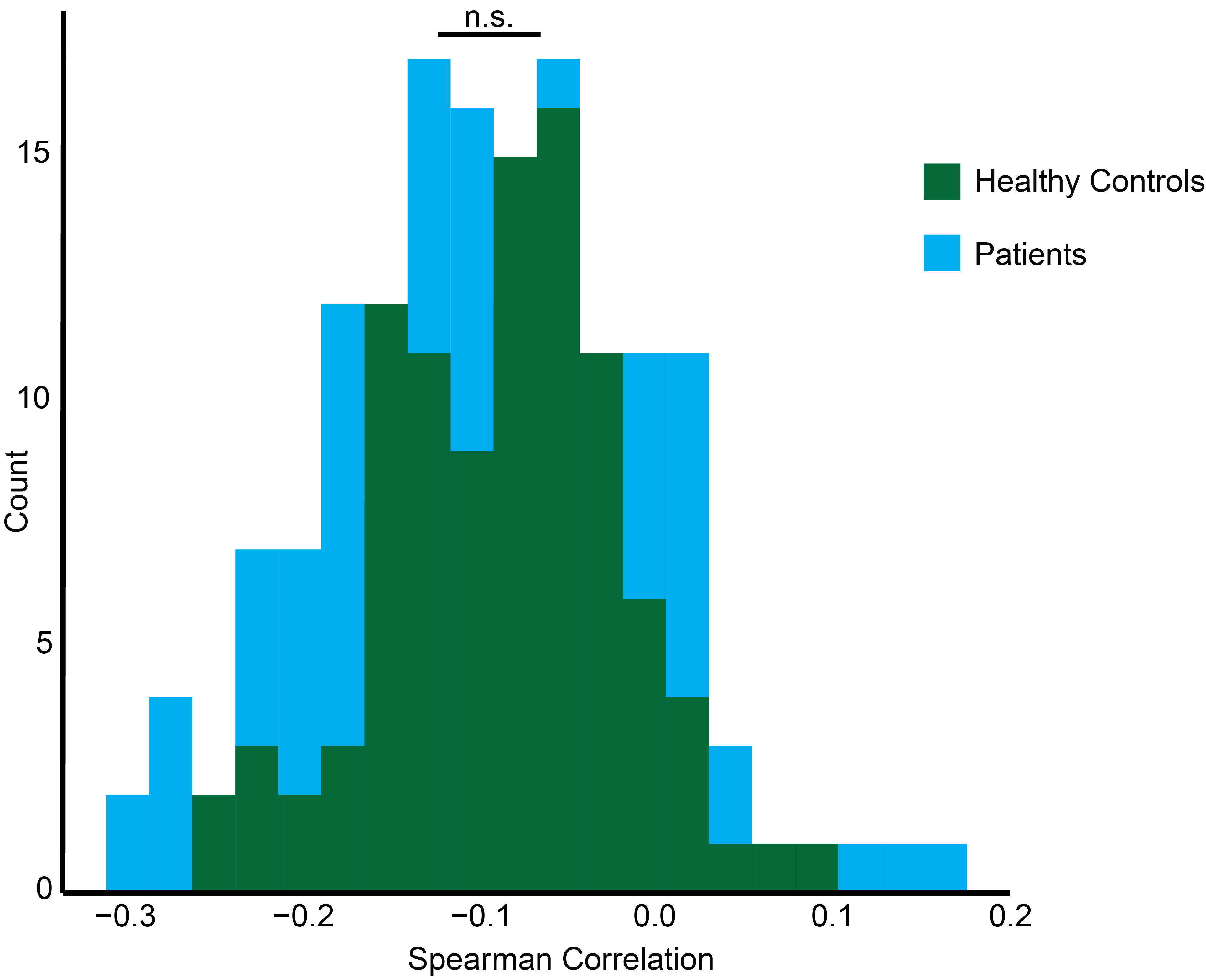}
	\caption{\textbf{Distributions of correlations across GM and SWM z-scores: } Comparison of correlations across modalities for patients and controls.}
	\label{Correlations}
\end{figure}

Finally, the correlation values between GM and SWM z-scores in patients did not significantly differ from those in controls (W = 7119, p = 0.6592) (Figure S8). This potentially means that the two modalities present complementary abnormalities that are contributing to epileptogenic activity. 

\section{Bootstrapping analysis on the individual ROC analyses}
\begin{figure}[H]
	\centering
	\includegraphics[scale=1]{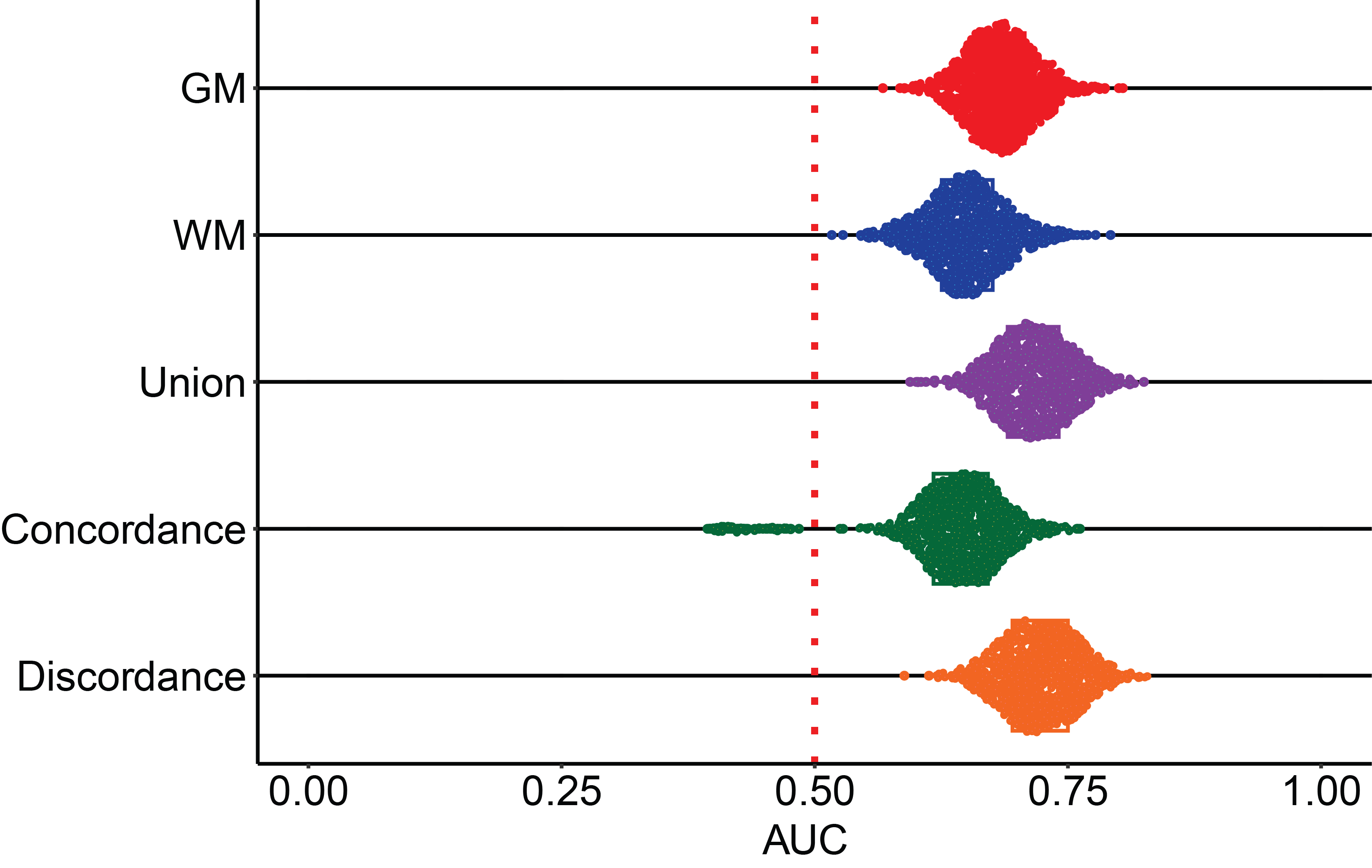}
	\caption{\textbf{Bootstrapping analysis on the individual ROC curves: } Distribution of AUC values based on the resampled datasets for GM, SWM, Union of GM and SWM, Concordance and Discordance across GM and SWM. Each dot is an individual AUC value based on the corresponding resampled dataset.}
	\label{Correlations}
\end{figure}

We conducted a bootstrapping analysis on the individual ROC curves, using 1,000 bootstrap samples with replacement, each maintaining the original sample size. This approach allowed us to assess the robustness of the AUC values in distinguishing between outcome groups. The results demonstrated that the AUCs remained consistent and stable across the resampled datasets, indicating the reliability of the model’s performance (Figure S9).

\vspace{5mm}

\section{Average proportion of abnormal regions and the proportion of resected abnormal regions}
\begin{center}
\begin{tabular}{ |p{8.9cm}|p{1cm}|p{1cm}|p{1.2cm}|p{2.2cm}|p{2.2cm}|} 
\hline
 & GM & WM  & Union & Concordance & Discordance \\
\hline
Mean proportion of regions identified as abnormal & 1.5\% & 3.8\% & 5.1\% & 0.2\% & 4.8\%  \\ 
\hline

\hline
Mean proportion of abnormal regions resected & 17\% & 18.2\% & 16.5\% & 33.8\% & 15.7\% \\ 
\hline

\end{tabular}
\end{center}

\vspace{5mm}

\section{Definition of terms}
\begin{center}
\begin{tabular}{ |p{4cm}|p{9cm}|} 
\hline
 Term & Definition \\
\hline
Union &  The union of two modalities (Grey Matter and Superficial White Matter) is the set that contains all the regions of interests that are deemed abnormal either in Grey Matter or Superficial White Matter, or both.\\ 
\hline
Concordance & The concordance of two modalities (Grey Matter and Superficial White Matter) is the set that contains all the regions of interests that are deemed abnormal in both modalities.\\ 
\hline
Discordance &  The discordance of two modalities (Grey Matter and Superficial White Matter) is the set that contains all the regions of interests that are deemed abnormal either in Grey Matter or Superficial White Matter, but not in both.\\ 
\hline

\end{tabular}
\end{center}

\end{document}